\label{fig:leading}       
\documentclass[epjCONF, onecolumn]{svjour}
\usepackage{graphicx}
\usepackage[varg]{txfonts} 
\usepackage[latin1]{inputenc}
\session-title{Assembling the puzzle of the Milky Way}
\begin{document}
\title{Towards understanding the dynamics of the bar/bulge 
       region in our Galaxy}
\author{E. Athanassoula\inst{1}\fnmsep\thanks{\email{lia@oamp.fr}}} 
\institute{Laboratoire d'Astrophysique de Marseille (LAM), UMR6110, 
CNRS/Universit\'e de Provence,\\
Technop\^ole de Marseille Etoile, 38 rue Fr\'ed\'eric Joliot Curie, 13388 Marseille C\'edex 20, France}
\abstract{
I review some of the work on bars which is closely linked to the
bar/bulge system in our Galaxy. Several independent studies, using
totally independent methods, come to the same results about the 3D
structure of a bar, i.e., that a bar is
composed of a vertically thick inner part and a vertically thin outer
part. I give examples of this from simulations and substantiate the discussion
with input from orbital structure analysis and from observations. The thick
part has a considerably shorter radial extent than the thin part. I then see
how this applies to our Galaxy, where two bars have been
reported, the COBE/DIRBE bar and the Long bar. Comparing their extents 
and making the reasonable and necessary assumption that our Galaxy has
properties similar to those 
of other galaxies of similar type, leads to the conclusion that these
two bars 
can not form a standard double bar system. I then discuss arguments in
favour of the two bars being simply different parts of the same bar,
the COBE/DIRBE bar being the thick inner part and the Long bar
being the thin outer part of this bar. I also very briefly discuss
some related new results. I first consider bar formation 
and evolution in disc galaxies with a gaseous component
-- including star 
formation, feedback and evolution -- and a triaxial halo. Then I
consider bar formation in a fully cosmological context using hydrodynamical
LCDM simulations, where the host galaxies grow, 
accrete matter and significantly evolve during the formation and
evolution of the bar. 
} 
\maketitle
\section{Introduction}
\label{intro}

Bars are present in roughly two thirds of disc galaxies in the local Universe,
so that the fact that our Galaxy is also barred is not exceptional.
Bars have also been observed in galaxies at higher redshifts and 
the fraction of disc galaxies having a bar decreases
with increasing redshift (\cite{Sheth} and references therein).

In this talk, I will will briefly review some of the theoretical 
results on bars which are relevant to our Galaxy and and apply them to
get a better understanding of the Milky Way bar/bulge system. I will
also make comparisons with observational results from
external galaxies. The latter is not always
straightforward, because both the viewing angle and the type of information
one can obtain in these two cases are considerably different. We are
located within 
the disc of our Galaxy, so that we have to rely on our experience with
face-on external galaxies for the face-on view. Furthermore, 
dust obscuration is a major problem, particularly in directions near
to that of the Galactic centre. On the other hand, it is possible
to observe in the Milky Way
individual stars in many directions and to obtain information on their
metallicity and line-of-sight velocity, which is, in general, still
not possible for galaxies outside our local group. In this context, the
most relevant dynamical studies are those concerning the vertical
bar instabilities, the corresponding orbital structure and the
growth of boxy/peanut bulges (hereafter b/p, or b/p bulges). 

I will also discuss two more bar-related studies. The
first one concerns some recent results on the effect of
gas and of triaxial haloes on the formation and properties of the bar
and peanut and the corresponding halo evolution. This work was done in
collaboration with Rubens Machado and Sergey Rodionov. The
second one concerns the formation of bars in fully 
cosmological hydrodynamical simulations, including comparisons of the
resulting bar properties 
to both those of dynamical (idealised) simulations and those of
observed bars. This work was done in collaboration with Cecilia
Scannapieco.  

\section{The 3D structure of bars}
\label{sec:3Dorbits}

\begin{figure}
\resizebox{0.50\columnwidth}{!}{\includegraphics{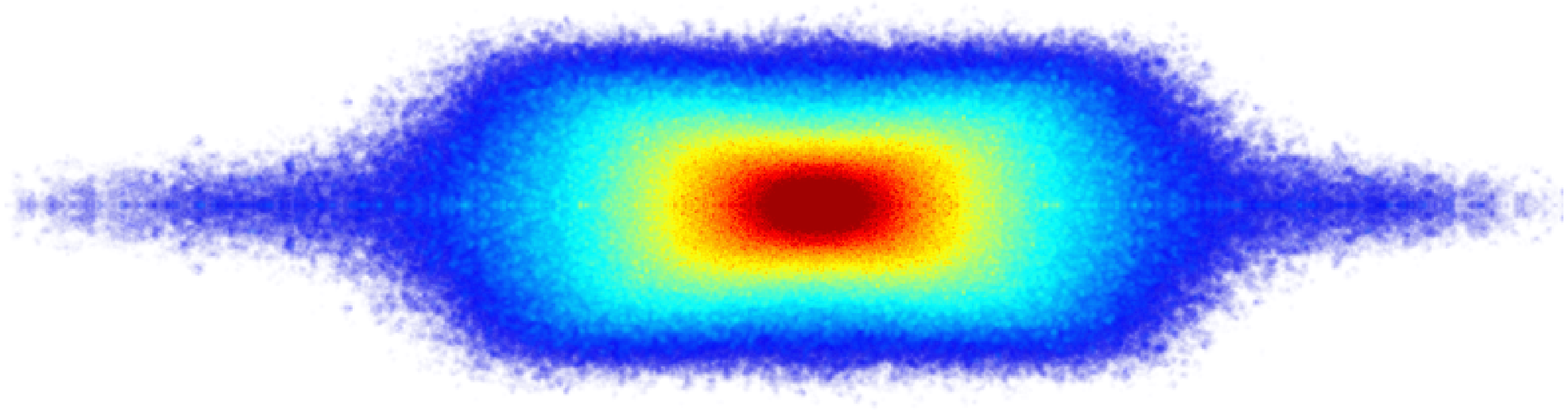} }
\resizebox{0.50\columnwidth}{!}{\includegraphics{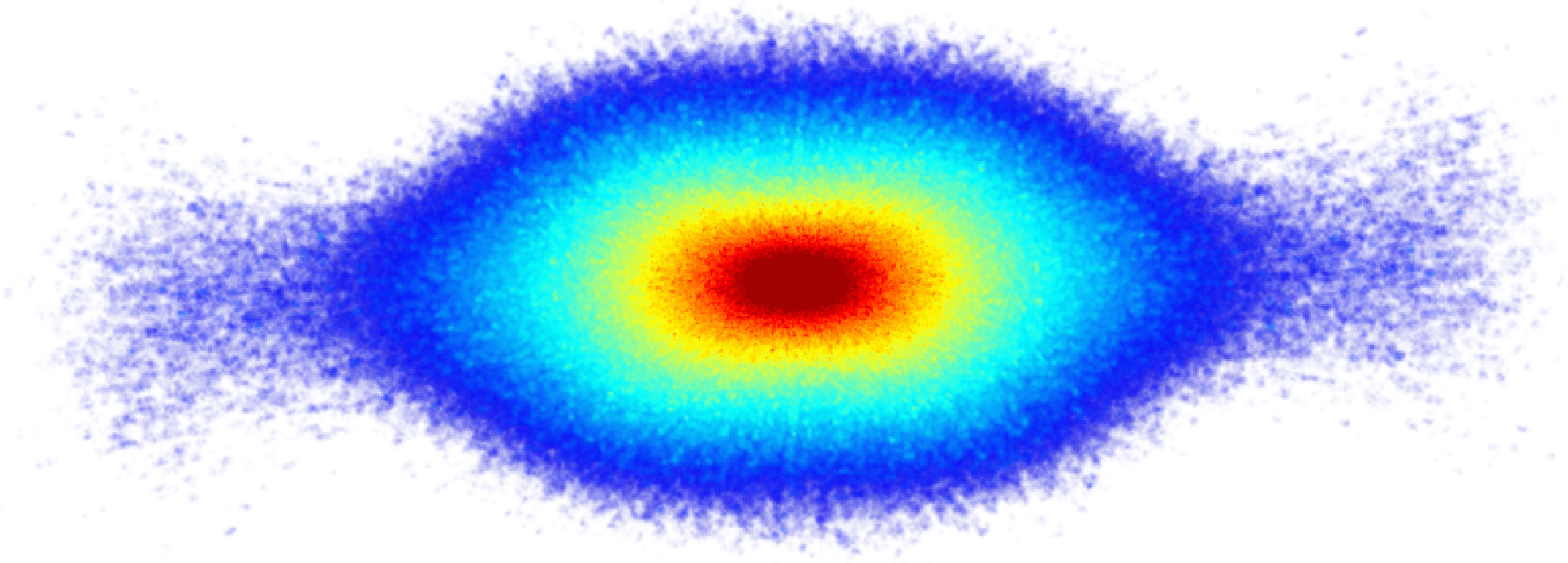} }
\caption{Side-on (left) and face-on (right) views of the bar component in a
  simulation. The side-on view shows clearly the boxy bulge (which is
  part of the bar) and the thin outer parts of the bar. The right panel
  view shows their face-on shape. These projections are produced with
  the help of the GLNEMO
  software (http://projets.oamp.fr/projects/glnemo2) and thus
  both are in 
  perspective and can not be used to calculate length ratios
  directly. See text for definition of the bar component.}  
\label{fig:2views-perspective}       
\end{figure}

Bars have a complex 3D structure. This is easy to see using N-body
simulations, where one can adopt any viewing angle one wishes. I will
therefore use such simulations to introduce and describe bar structure and
morphology. An example of an N-body bar seen 
side-on\footnote{In a side-on view the galaxy is seen edge-on and the
  line of sight is along the bar minor axis.} and face-on
is shown in Fig.~\ref{fig:2views-perspective}. I display here
only the bar component, neglecting the halo, the remaining disc and the ring
which were present in this simulation. It is of course always somewhat
arbitrary to decide which component a given particle
belongs to, because all particles move in the same potential, that of
the Galaxy. My
picking out only the bar particles is therefore also somewhat
arbitrary and I could have picked out more particles than I should
have had, or, more likely the case here, neglected some particles which I could
have included. Nevertheless, Fig.~\ref{fig:2views-perspective}
shows very clearly the complex 3D structure of the bar. Over a
considerable fraction, but not all, of its 
extent, the bar is vertically thick, much thicker than the underlying disc. Its
outer parts, however, are quite thin. The face-on view shows clearly
that the
thick b/p component and the thin outer parts have the same position
angle\footnote{I
  define here the bar position angle in our Galaxy as the angle
  between the bar major axis and 
  the line from the centre of the Galaxy to the Sun.} and that the
latter, at least in this example, has the form of handles (ansae). It
also shows clearly that the thick and the thin parts of 
the bar have different radial extents, as has been already noted in a number
of simulations (see \cite{ath05} for a summary and discussion).

\begin{figure}[h!]
\begin{minipage}{0.45\columnwidth}
\includegraphics{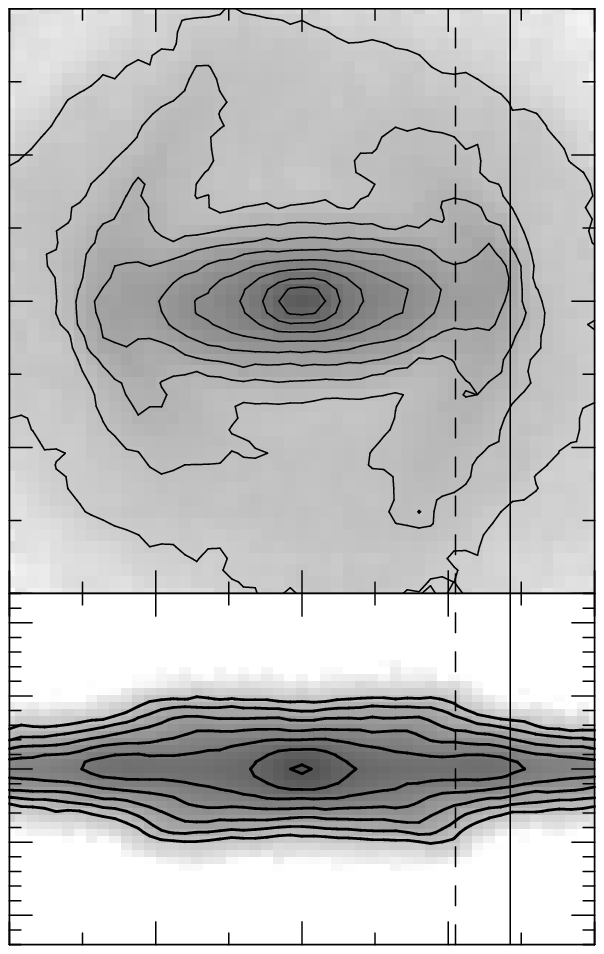}
\label{fig:3views-orthogonal}       
\end{minipage}
\begin{minipage}{0.55\columnwidth}
{\bf Fig. 2.} Face-on (top) and side-on (bottom) views of all disc
  particles in a N-body simulation 
  having formed a strong bar. The solid and dashed vertical lines give
  estimates of the bar length and of the peanut length,
  respectively. The projection, contrary to Fig. 1,
is orthogonal. 
\vskip 0.65truecm

\normalsize
provides building
blocks both for the bar and the b/p. The 
$x_1$ orbits are the backbone of the bar, while one or more of the
vertical families provide building blocks for the b/p. The extent
along the bar major 
axis is different in the two cases and orbital structure theory shows
that the extent of the family providing building blocks 
for the b/p is considerably shorter than that of the family producing
the bar. In other words, the
b/p should be considerably {\it shorter} than the bar, the ratio of
the two extents depending on which of the vertical families is the
crucial contributor to the b/p
(see \cite{Patsis.SA}, where a number of examples and length ratios are given),
in good agreement with what N-body simulations show.

~~~~~~~Photometry of edge-on galaxies including cuts along their major axis, as
well as
parallel and offset from it (\cite{ldp00b,Bureau}) also argue in
favour of a b/p which is shorter than the bar. This bar structure has
also been studied in galaxies not edge-on but not far from it 
(e.g. \cite{BetGal,Quillen+,Athanassoula.Beaton.06}).

\end{minipage} 
\end{figure}

A more quantitative measure of the difference between the bar and peanut
lengths is given in
Fig.~\ref{fig:3views-orthogonal}. Here I show the disc component of an
N-body simulation, or to be more precise, its bar region,  after the bar
has grown. From the
upper panel (face-on view) it is possible to get a rough estimate of
the bar length and from the bottom panel (side-on view) a rough
estimate of the b/p length. They are indicated by a solid and a dashed
vertical line, respectively. Their location shows clearly that the bar is
considerably more extended than the b/p bulge. Let us now see how this 
clear, albeit complex, picture compares with the results from orbital
structure and from observations.

In the simple 2D picture, the building blocks for bars are provided by
a family of periodic orbits, elongated along the bar and called $x_1$ 
(\cite{Contopoulos.Papayannopoulos.80,Athanassoula.BMP.83}).
In 3D, however, the structure is more complex and the building blocks of this
bar and of its boxy/peanut thick part come from a tree of 2D families as well
as from 3D families bifurcating from the $x_1$ family (\cite{Pfenniger,sko02a}).
However, it is not the same family that 

 Thus all three 
approaches -- orbital structure, N-body simulations and observations --
argue cohesively that the length of the b/p is less than that of
the bar. I will now discuss how this can influence our view of the
Milky Way's bar/bulge system. 
   
\section{The Milky Way bar/bulge system}
\label{sec:MW}

Considerable amount of evidence has accumulated since the 1990s to
argue that our Galaxy is barred
(e.g. \cite{wei94,Dwek,BineyGS,Stanek,Dehnen,LC,Hamadache,Minchev}). This
bar, seen clearly in the NIR images, is 
often referred to as the COBE/DIRBE 
bar. More recently, further evidence coming from
star counts at larger longitudes opened the possibility for a second
bar, longer than the first one and at a larger position angle
(e.g. \cite{ham00,Benjamin,CL07,lop07,CL08,Churchwell}).    

The existence of a second bar is not uncommon. About a fourth or
a fifth of disc galaxies have both a primary, or main bar and a
secondary, or inner bar (\cite{ErwinS,Laine.SKP02,Erwin09}). 
Secondary bars have lengths between 0.1 and 
1.2 kpc, and in relative size they are about 10 - 15\% of the main
bar. These numbers contrast strongly with those found for our Galaxy,
where the length of the COBE/DIRBE bar is found to be about 3 - 3.5 kpc
and that of the Long bar about 4 kpc. Double bars have also been found
in N-body simulations of disc galaxies, with relative lengths that
agree well with observations, and not with those of the COBE and Long
bar in the MW (\cite{Heller.SA07,Shen.Debattista.09}).
It can thus be excluded that the
COBE/DIRBE and the Long bar are a double bar system similar to that
seen in external galaxies. They could indeed form a double bar system
only if this was very different from those observed in external
galaxies. Since it is hazardous to assume that our Galaxy is very
different morphologically from external ones, this
possibility should be discarded.    

So, if the COBE/DIRBE bar and the Long bar do not constitute a double
bar system, then what are they?
The data give us two important clues on that. First, the ratio of the
major-- to $z$-- semi-axis of the bar is $\sim$0.3 for the COBE/DIRBE
bar and $\sim$0.026 for the Long bar, i.e. {\it the Long bar is very thin
and the COBE/DIRBE bar is very thick}. Second, {\it the ratio of the
lengths of these two bars ($\sim$0.8) is in good agreement with the
ratio of the extent of the b/p to that of the thin outer part of bars
found by orbital structure analysis and N-body simulations} (e.g. 
\cite{Patsis.SA,Athanassoula.Beaton.06}). Both these clues point to the fact 
that there is only one bar in our Galaxy of which the thick
inner part (b/p bulge) is the COBE/DIRBE bar and the thin outer part
is the Long bar. This bar structure, in good agreement with theory and
observations from external galaxies, was 
first proposed for our Galaxy by \cite{Athanassoula08}
and first tested by \cite{CL07} using their red clump
giants measurements.

\begin{figure}[ht!]
\begin{minipage}{0.55\columnwidth}
\includegraphics[width=70mm]{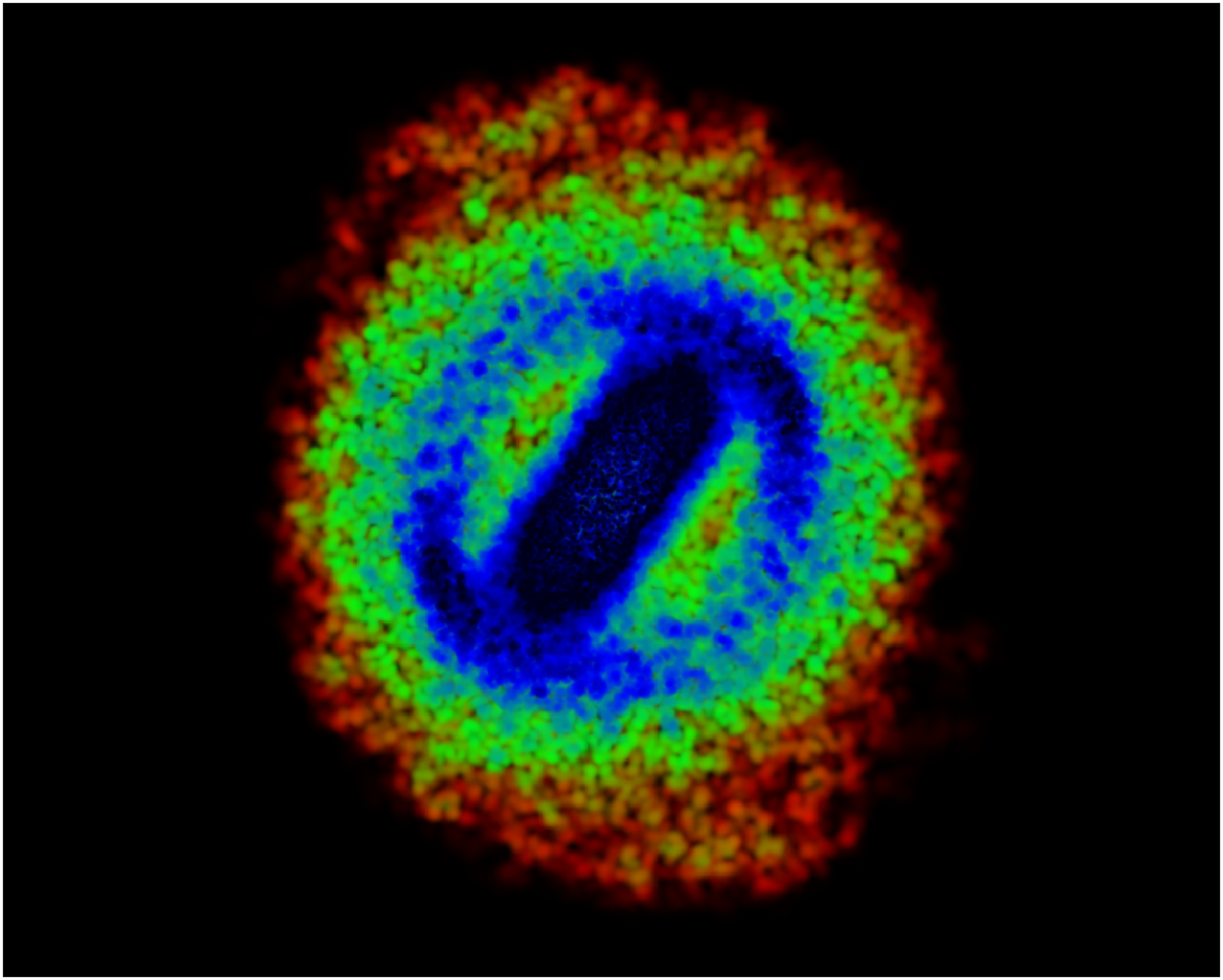} 
\label{fig:leading}
\end{minipage}
\begin{minipage}{0.45\columnwidth}
{\bf Fig. 3.} Face-on view of the disc component in a simulation with a
  strong bar and a ring. Rotation here is clockwise. Note that
  the density along the ring is not constant, but is in fact somewhat
  higher in the leading than in the trailing side. If such a structure
  is also present in our Galaxy, it could result in an estimated bar position
  angle somewhat larger than the actual value.
\end{minipage}
\end{figure}
 
This would have been a clear and undisputed result, had it not been
for the fact that observations argue for the the COBE/DIRBE bar and the
Long bar having considerably different position angles, between 15$^{\circ}$
and 30$^{\circ}$ for the former and around 43$^{\circ}$ for the latter.
How can this be reconciled with the single bar picture
presented above? Let me first say that, due to our location within the
Galaxy (see Sect. 1), the estimates for the Galactic bar position
angles are much less 
accurate than the corresponding estimates for external galaxies. Thus
Zasowski, Benjamin \& Majewski (poster at this meeting) find the
position angle of the Long bar to be between 25$^\circ$ and 35$^\circ$,
i.e. much closer to that of the COBE/DIRBE bar than the 43$^{\circ}$
estimated by \cite{ham00,Benjamin,CL07,CL08}.

A further point to note is that, if the outer isodensity contours of
the thin part of the bar are, in the equatorial plane, more
rectangular-like than elliptical-like, as is often the case in
external galaxies (e.g. \cite{AthaMWPPLB90,Gadotti08,Gadotti10}), the bar
position angle will appear to be at larger angles than what it
actually is. A final important point is that in many N-body simulations
there is often an inner ring and within it a short, leading segment
near the end of the bar (see
e.g. \cite{Athanassoula.Misiriotis.02,mar11,rom11}, and Fig.~\ref{fig:leading}
here). In view of all the 
above comments, the small difference in position angle between the
COBE/DIRBE and the Long bar, should not be a major concern.

We are thus led to the conclusion that the COBE/DIRBE and the Long bar
are 
parts of the same bar, and do not constitute two separate components. 
There is, therefore, good agreement between, on the one hand, the properties
of the bar/bulge system in our Galaxy, and, on the other, observations
of external galaxies, simulations
and orbital structure theory. Such an agreement is of
paramount importance, since our Galaxy should not be different from other disc
galaxies of the same type.

A more extended version of the discussion presented in sections 2 and
3 can be found in \cite{rom11}. 
After my talk, I. Martinez-Valpuesta presented a contribution reaching
a very similar result, although in a different way. This work is now
published (\cite{mar11}). 

\section{Further results on bars}
\label{sec:more-results}

\subsection{Bars formed in disc galaxies with gas and with triaxial haloes}
\label{sec:rubens}

Although quite extensive, most previous studies of bars rely on at least one
of the two following simplifying approximations: That haloes are
spherical and that the gaseous content of disc galaxies can be
neglected, or at least modelled in an oversimplified way. Both these 
approximations were abandoned by Athanassoula, Machado \& Rodionov (in
prep.), who considered triaxial haloes and a gaseous content of the
disc modelled including star formation, feedback and cooling.
They find that these new effects considerably influence the bar
morphology, strength and length, as well as the gas distribution. In
general, haloes evolve towards axisymmetry even if initially very
triaxial, or in the presence of a considerable gaseous component. 
In several previous studies, albeit with spherical haloes and no gas, it was
shown that the inner part of the halo becomes prolate, forming what
was termed a `halo bar', or `dark bar' (e.g. 
\cite{Atha05b,Colin+2006,BerentzenShlosman2006,Ath07,Machado.Athanassoula.10}).
Our study confirmed these 
results for cases where the initial gaseous content is rather low and
the halo either spherical, or mildly triaxial. In the remaining cases,
however, we find that the central regions become, on the contrary,
more spherical.    
  
\subsection{Bars in a cosmological framework}
\label{sec:cosmo}

Scannapieco \& Athanassoula (in prep.) discuss two examples of the
formation of bars in fully cosmological 
hydrodynamical simulations and compare the resulting bar properties
to both those of bars formed in dynamical (idealised) simulations and
to those of 
observed bars. Some of their results can be found in an upcoming
paper and show that realistic bars can form
naturally in $\Lambda$CDM simulations, where host galaxies grow,
accrete matter and significantly evolve during the formation and
evolution of the bar. More specifically, this work addresses the
length and strength of bars and their effect on the radial projected
density profiles and on the velocity field. 

\par\vspace{5mm}

{\bf Acknowledgements} I thank  my collaborators Albert Bosma,
Merce Romero-Gomez, Rubens Machado, Sergey Rodionov and Cecilia
Scannapieco for interesting and 
motivating discussions, Jean-Charles Lambert for his help with GLNEMO
and the organisers for inviting me to this very interesting meeting.

\end{document}